\RequirePackage{lineno}
\documentclass[prl,showpacs,amsmath,amssymb,twocolumn,superscriptaddress]{revtex4}

\usepackage{graphicx}
\usepackage{graphics}
\usepackage{dcolumn}
\usepackage{bm}
\usepackage{amsmath}
\usepackage{url}

\begin{document}
\preprint{}
\title{Production of Ultra-Cold-Neutrons in Solid $\alpha$-Oxygen.}

\author{E. Gutsmiedl$^{*}$}
\affiliation{Technische Universit\"at M\"unchen,Physik Department,
James Franck Str. 1, D-85747 Garching, Germany}

\author{F. B\"ohle}
\affiliation{Technische Universit\"at M\"unchen,Physik Department,
James Franck Str. 1, D-85747 Garching, Germany}

\author{A. Frei}
\affiliation{Technische Universit\"at M\"unchen,Physik Department,
James Franck Str. 1, D-85747 Garching, Germany}

\author{A. Maier}
\affiliation{Technische Universit\"at M\"unchen,Physik Department,
James Franck Str. 1, D-85747 Garching, Germany}

\author{S. Paul}
\affiliation{Technische Universit\"at M\"unchen,Physik Department,
James Franck Str. 1, D-85747 Garching, Germany}

\author{H. Schober }
\affiliation{Institut Laue Langevin, 6 rue Jules Horrowitz,
F-38042 Grenoble Cedex 9, France} \affiliation{Universite Joseph
Fourier, UFR de Physique, F-38042 Grenoble Cedex 9, France}

\author{A. Orecchini }
\affiliation{Institut Laue Langevin, 6 rue Jules Horrowitz,
F-38042 Grenoble Cedex 9, France} \affiliation{Dipartimento di
Fisica, Università di Perugia, I-06123 Perugia, and CNR-INFM
CRS-Soft c/o Università di Roma "La Sapienza", I-00185 Roma,
Italy.}

\date{\today}

\begin{abstract}

Our recent neutron scattering measurements of phonons and magnons
in solid $\alpha$-oxygen have led us to a new understanding of the
production mechanism of ultra-cold-neutrons (UCN) in this
super-thermal converter. The UCN production in solid
$\alpha$-oxygen is dominated by the excitation of phonons. The
contribution of magnons to UCN production becomes only slightly
important  above $E>$10~meV and at $E\sim$4~meV. Solid
$\alpha$-oxygen is in comparison to solid deuterium less efficient
in the down-scattering of thermal or cold neutrons into the UCN
energy regime.
\end{abstract}


\noindent
$^*$Corresponding author; email: egutsmie@e18.physik.tu-muenchen.de\\

\maketitle

Ultra-cold-neutrons (UCN) are slow enough ($\sim$~300~neV) to be
confined \cite{Golub1} in traps, which can be formed by materials
with a high Fermi potential or by a magnetic field (60 neV/T).
They can be kept for several minutes in the confinement, and thus
be investigated with high precision. UCN are elementary particles
that are extremely well suited for low- energy physics
experiments. These experiments are investigating fundamental
problems unsolved within the framework of the Standard Model
\cite{StanM}. One major experiment is the search for a non-zero
electric dipole moment of the neutron \cite{Bak} (current upper
limit 2.9~10$^{-26}$ e$\cdot$ cm). Another unique experiment is
the precise determination of the lifetime \cite{Paul} of the free
neutron. This value has an important impact on the theory of weak
interaction \cite{Schreckenbach}.
\\

\begin{figure}[htp]

\vspace{-2mm}

\includegraphics[width=0.5\textwidth]{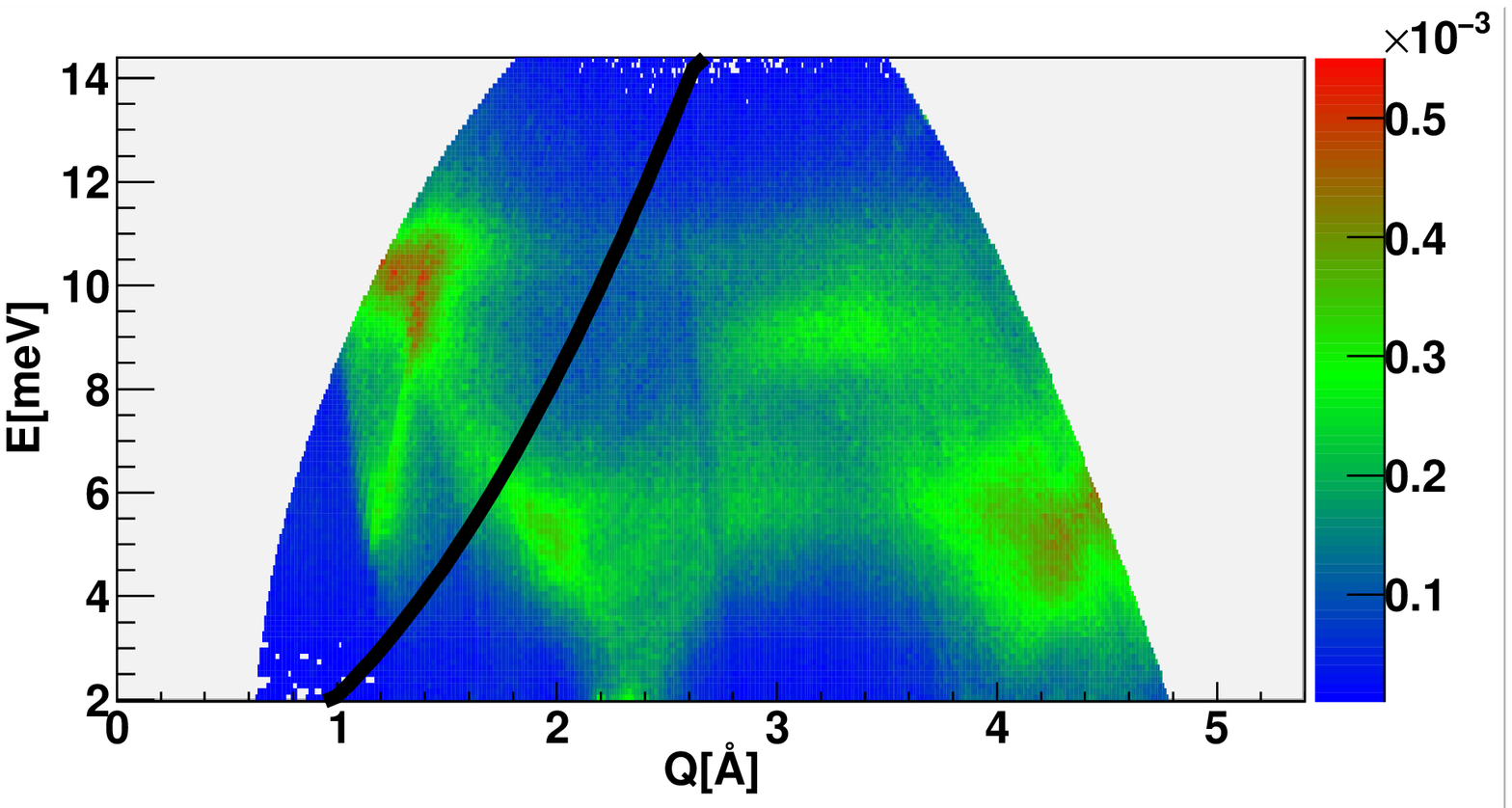}


\caption{~}{$S(Q,E)$ (arib. units) of $\alpha$-sO$_2$ at $T=5$~K.
Data from IN4 measurements. Black parabola: Dispersion of the free
neutron}

\label{fig.1}


\end{figure}
Powerful UCN sources are needed for the experiments mentioned
above in order to minimize the statistical errors, and different
groups \cite{Frei-1,LANL,PSI-1,SER} are working on the development
of strong UCN sources, based on solid deuterium (sD$_2$) as a
converter for down-scattering of thermal or sub-thermal neutrons
into the UCN energy region. Solid oxygen could be a valuable
alternative when grown in the $\alpha$-phase ($\alpha$-sO$_2$).
Solid $\alpha$-sO$_2$ has a 2-dimensional anti-ferromagnetic
structure \cite{AFM}, which exhibits in addition to phonons spin
wave excitations (magnons). This supplementary magnetic scattering
of neutrons considered for the first time by {\it{ Liu and Young}}
\cite{INDI} might be a strong down-conversion channel, which would
enhance the production of UCN. Different groups
\cite{PSIIND-1,PSIIND-2} performed experiments concerning UCN
production in such a converter. Their results are inconclusive and
a challenge to investigate $\alpha$-sO$_2$ further. It seems that
preparation of this cryo-solid is crucial. The exact knowledge of
the inelastic scattering channels in solid $\alpha$-O$_2$ is
therefore very important.
\\
To this aim we have measured the phonon/magnon system in
$\alpha$-sO$_2$ by neutron time-of-flight (TOF) measurements at
the IN4 spectrometer (Institute Laue-Langevin Grenoble - ILL).
Thermal neutrons with an energy of $E_0 =16.7$~meV were used to
determine the scattering function $S(Q,E)$ in the range from
$0-14$~meV. The experimental setup ( sample cell, gas system and
slow control) was the same as it was used in the measurements of
the dynamical neutron scattering function of sD$_2$
\cite{UCNPsD2}. We used oxygen gas with a purity
$\geq$~99.999$\%$. Our measurements were performed without any
external magnetic field. The $\alpha$-sO$_2$ crystals were
prepared from liquid via the $\gamma$- and $\beta$-phase
\cite{Freiman}. The phase transition $\gamma$ to $\beta$ at T=43.8
~K at vapor pressure was done in our experiments very slow
(10~mK/h) in order to get optical semi-transparent crystals
\cite{Raliza}.

\begin{figure}[htp]
\includegraphics[width=0.5\textwidth]{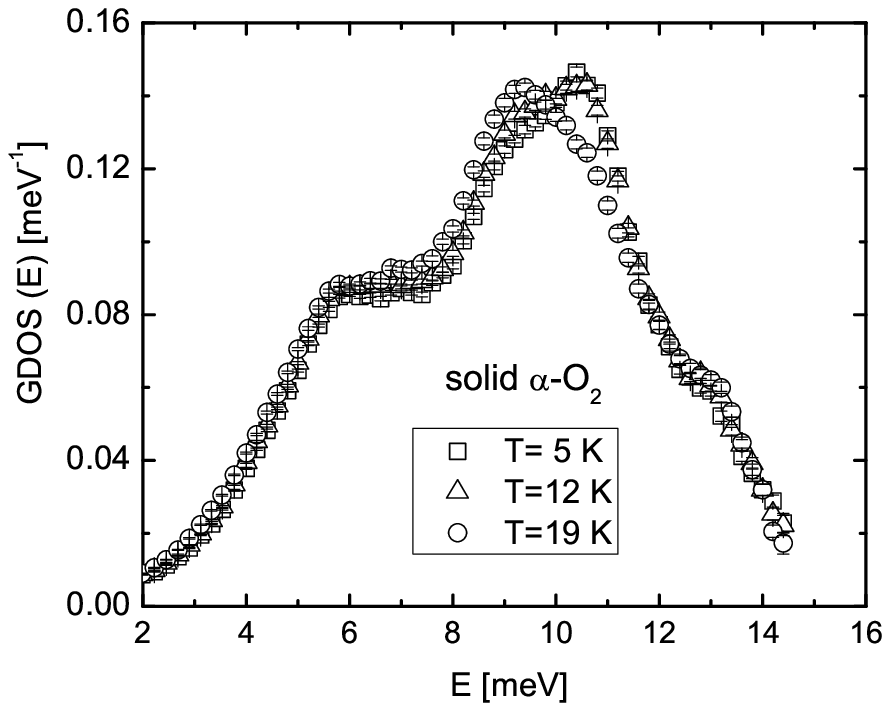}
\caption{~}{Generalized density of states $GDOS(E)$ of
$\alpha$-sO$_2$ at
5~K~($\square$)/12~K~($\bigtriangleup$)/19~K~($\bigcirc$). Data
from IN4 measurements. GDOS is normalized to
$\int_0^\infty{GDOS(E)\cdot dE}=1$} \label{fig.2}
\end{figure}

\begin{figure}[htp]
\includegraphics[width=0.5\textwidth]{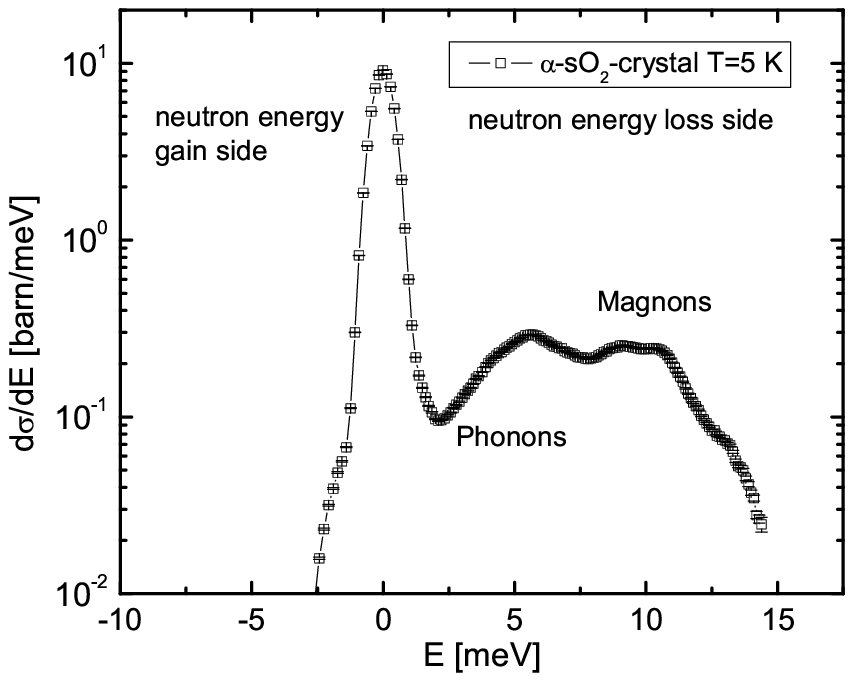}
\caption{~}{$d\sigma /dE$ of $\alpha$-sO$_2$ at $T=5$~K. Data from
IN4 measurements.} \label{fig.3}
\end{figure}

The measured scattering function $S_{\rm{data}}(Q,E)$, the
generalized density of states GDOS(E) and the cross section
$d\sigma /dE$ of $\alpha$-sO$_2$ are shown in Fig.\ \ref{fig.1},
Fig.\ \ref{fig.2} and Fig.\ \ref{fig.3}. Theoretical calculations
predict \cite{INDI, Jansen} magnetic excitations at E$\sim
1-4$~meV. The intensities of these excitations should be primarily
found at low Q-values ($Q\ll$~1~$\mathrm{\AA^{-1}}$). They thus do
not fall into the observation window of our experiment on the
energy-loss side (see Fig.~\ref{fig.1}). However, the free neutron
parabola does not cross this region and as consequence those
excitations cannot contribute to the down-scattering process from
the thermal into the UCN-region (see further down). The cross
section d$\sigma$/dE (see Fig.~\ref{fig.3}) shows on the energy
gain side only up-scattering close to E=0 (elastic peak). This
up-scattering in our data is very likely trigged by phonons. This
experimental observation should induce a large mean free path
($\lambda _{mfp}$) of UCN in $\alpha$-sO$_2$, as it was predicted
by {\it{ Liu and Young}} \cite{INDI}.
 Neutron scattering by solid oxygen is purely coherent and mostly elastic
($\sigma _{el}/\sigma _{tot}\sim$ 0.84 - value deduced from our
neutron scattering data - see Fig.~\ref{fig.3}). The elastic Bragg
peaks in Fig.~1 are cut out in order to enhance the contrast for
the inelastic scattering in the plot.
\\
The GDOS can be calculated from $S(Q,E)$ \cite{Squires} by
sampling over a large $Q$-range (neutron energy loss side), and
applying the incoherent approximation \cite{TURCH}. The GDOS can
be deduced by

\begin{equation}
\label{eq.1}
\begin{aligned}
GDOS(E) = E \cdot\ \frac{S(Q(<2\theta>),E)} {Q^2 \cdot (n+1)}.
\end{aligned}
\end{equation}

The term $Q(<2\theta>)$ is a Q value for each energy transfer $E$
considering an averaged scattering angle $<2\theta>$, while $n$ is
the Bose distribution for the phonons.

Although oxygen is a purely coherent scatterer, it can be useful
to use this approximation to obtain an estimate of the excitation
spectra in the sample \cite{Breuer}.  First results of the
generalized density of states (T$\simeq$4~K and 23~K) were
published by {\it{de Bernabe et al.}} \cite{Bernabe} and
{\it{Kilburn et al.}} \cite{Kilburn}. Their results show a mixture
of phonons, librons and anti-ferromagnetic excitations (magnons).
The peak at E$\simeq$10.5~meV at T=5~K in our result is more
pronounced compared to the result of {\it{de Bernabe et al.}}. The
GDOS  at T=10~K of {\it{Kilburn et al.}} is more equal to our
result at T=5~K. At higher temperatures our GDOS and {\it{de
Bernabe et al.}} GDOS are showing similar structures. Calculated
contributions of magnons (see Fig.~6 in \cite{Bernabe}) should
appear at E$\simeq$5~meV and E$\simeq$12.5~meV, which are neither
detected in our or their data. The  magnon peak positions at lower
energies are explained by {\it{de Bernabe et al.}} by a decrease
of the exchange constant (see Eq.~(6) in \cite{Bernabe}) with
decreasing temperature. A more general detailed analysis of our
neutron scattering data will be presented in a forthcoming paper.

These experimental findings in our data have an important impact
on the UCN production in solid oxygen. The dynamical scattering
function of solid oxygen resolved from our neutron scattering data
has to be calibrated to absolute values. This calibration uses the
known value of the total cross section for thermal neutron
energies.

\begin{equation}
\label{eq.2}
 \sigma _{\rm{tot}}(E_0)={\int_{0}^{\infty}
dE_{\rm{f}} \int_{}^{} \frac {k_{\rm{f}}}{{\it k_0}} {{\it
b_{\mathrm{eff}}}}^{2}S\left(Q,E \right) d\Omega}.
\end{equation}

\begin{figure}[htp]
\includegraphics[width=0.5\textwidth]{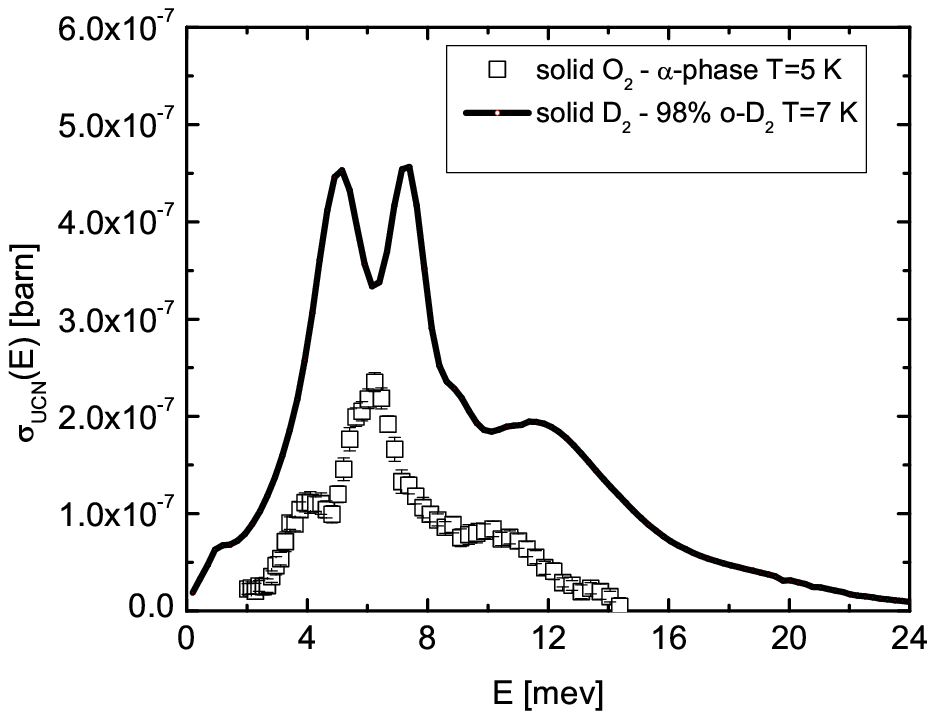}
\caption{~}{UCN production cross section of $\alpha$-sO$_2$ (
$\square$) at $T=$5~K and sD$_2$ (98$\%$ ortho concentration) (
line) at $T=$~7~K. UCN energy range 0-140 neV inside the solid
D$_2$, UCN energy range 0-163 neV inside the solid
$\alpha$-sO$_2$. Cross section determined by a integration of
$S(Q,E)$ along the free dispersion of the neutron. Data from IN4
measurements.}

\label{fig.4}
\end{figure}

The wave vector and the energy of the scattered neutrons in Eq.\
\eqref{eq.2} are $k_{\rm{f}}$ and $E_{\rm{f}}$, whereas $k_0$ is
the wave vector of the incident neutrons. The effective scattering
length $ b_{\rm{eff}}^2 = 2\cdot b_{\rm{nucl}}^2+b_{\rm{mag}}^2$
contains a combination of nuclear ($b_{\rm{nucl}} = 5.8$~fm
\cite{NIST}) and magnetic scattering ($b_{\rm{mag}} = 5.38$~fm
\cite{INDI}). The dynamical scattering function can be calculated
via

\begin{equation}
\label{eq.3}
 S(Q,E)=\kappa \cdot S_{\rm{data}}(Q,E).
\end{equation}

The value of the calibration factor $\kappa=$1240 is obtained by
using Eq.\ \eqref{eq.2} and the knowledge of the total cross
section ($\sigma _{\rm{tot}}(E_0$=16.7~meV)$\approx 4\pi
b_{\rm{eff}}^2$=12.1~barn). In the case of UCN production the
following relations are valid: $E_f=E_U \ll E_0$; $E=E_0-E_f \sim
E_0$, where $E_0$ is the initial energy of the neutron before
scattering. The UCN production cross-section can determined by
direct integration of the dynamic neutron cross section data in
the kinematic region along the free neutron dispersion parabola
($E_0\approx =\hbar^2Q^2/2m$)

\begin{equation}
\label{eq.4}
 \sigma_{\rm{UCN}}(E_0)=\int_{0}^{E_{\rm{U}}^{\rm{max}}} \frac {d\sigma
(E_{\rm U})} {dE_0} dE_{\rm{U}}.
\end{equation}

The evaluation of the integral (Eq.\ \eqref{eq.4}) uses the
dynamic scattering function $S(Q,E=\frac {\hbar^2} {2m} k_0^2)$ at
the phase space points of the neutron parabola. The UCN production
cross section can, therefore be expressed by

\begin{equation}
\label{eq.5}
 \sigma _{\rm{UCN}}(E_0)=\frac {\sigma _0} {k_0}
S(k_0,\frac {\hbar^2} {2m} k_0^2) \frac {2} {3}
k_{\rm{U}}^{{\rm{max}}} E_{{\rm{U}}}^{{\rm{max}}} .
\end{equation}

The term $E_0=\frac {\hbar^2} {2m} k_0^2$ is the energy for an
incoming neutron with wave vector $k_0$, whereas $\sigma _0$ is
the total cross section ($\sigma _0=4\pi b_{\rm{eff}}^2$). The
result for $\sigma _{\rm{UCN}}(E)$ is shown in Fig.~\ref{fig.4}.
For comparison the UCN production cross section for ortho-sD$_2$
is also included in Fig.~\ref{fig.4} (line).

When determining the upper limit of the integration we have to
take into account that the UCN will gain kinetic energy  when
leaving the converter \cite{PSI-2}. UCN produced in sD$_2$ gain
$\Delta E_{\rm{U}}\sim 110$~neV, while the energy gain for
$\alpha$-sO$_2$ is $\Delta E_{\rm{U}}\sim 87$~neV. Therefore, the
upper limit of the energy the neutrons are allowed to have inside
the converters was set to $E_U^{\rm{max}}(\alpha$-sO$_2)=$163~neV
and $E_U^{\rm{max}}($ortho-s$D_2)=$140~neV. These limits
  correspond to an upper limit of
$E_{\rm{U}}^{\rm{max}} =250$~neV outside the converter (Fermi
potential of UCN guide). The calculation of UCN production cross
section of sD$_2$ was performed  using the incoherent
approximation \cite{TURCH} and recently published data
\cite{UCNPsD2} for the density of states in sD$_2$.

In $\alpha$-sO$_2$ the parabola of the free neutron crosses in the
scattering function a band of dispersive excitations at $E\sim$ 6
meV (see Fig.~\ref{fig.1}). At this point  the UCN production
cross section is determined by coherent phonon scattering.
Therefore a major peak (Fig.~\ref{fig.4}) at $E\sim 6$ meV can be
 identified with excitation of phonons ( see also \cite{Jansen}).
The structures in the UCN production cross section at $E\sim
4$~meV and $E\sim 10$~meV are very likely induced by magnetic
scattering (magnons) \cite{Bernabe}. The contribution of magnons
at $E\sim 4$~meV and 10-12~meV to the UCN production cross section
is small compared to the phonon contribution at $E\sim 6$~meV. Our
data clearly lead to the conclusion that the creation of phonons
is the main energy loss channel in the conversion process of
$\alpha$-sO$_2$. In comparison to solid ortho-deuterium (see
Fig.~\ref{fig.4}) $\alpha$-sO$_2$ possesses a remarkably poorer
capacity of creating UCN by down-scattering of thermal and
sub-thermal neutrons. This result can be  explained by a larger
inelastic cross section of ortho-sD$_2$ ($T\simeq$7~K) compared to
$\alpha$-sO$_2$ at thermal neutron energies \cite{INWQ}. The ratio
of the two inelastic cross sections is $\sigma
_{\rm{inel}}$(ortho-sD$_2$)$/
\sigma_{\rm{inel}}$($\alpha$-sO$_2$)$\simeq$4.7 at $E_0
=16.7$~meV. On the other hand $\alpha$-sO$_2$ should exhibit a
large mean free path for UCN inside the converter as it is
predicted by {\it{ Liu and Young}} \cite{INDI} and also indicated
by our data (see Fig.~\ref{fig.3}). Values up to $\lambda
_{mfp}\simeq$~4~m are expected. This opens the opportunity to
construct a large UCN source with this material. This kind of
large source could defeat a sD$_2$ source due to the small mean
free path (several cm \cite{Frei-1}) of UCN in solid deuterium. We
recently have measured directly the UCN production in
$\alpha$-sO$_2$ \cite{Frei-2}. From our data we could extract a
$\lambda _{\rm{mfp}}\simeq$~0.3~m at $T\simeq$8~K. This mean free
path increases to $\lambda _{\rm{mfp}}\simeq$~3~m at $T\simeq$5~K
and confirms the predictions made by {\it{ Liu and Young}}
\cite{INDI}. Furthermore we could deduce an average final energy
of $E_f\sim$~3~ meV after up-scattering of UCN from our new data
\cite{Frei-2}. This value for the final energy could be provided
by the predicted magon excitations. At 5 K the occupation of such
moded is such that the corresponding intensities become
unobservable in our scattering experiment, while it is still
sufficient to provide observable upscattering of UCN.

The UCN production rate $P_{UCN}$~[UCN/$cm^3$~s] of
$\alpha$-sO$_2$, which is exposed to a neutron spectrum $d\Phi/dE$
with maxwellian energy distribution ($T_n$~- Effective temperature
of the incoming neutron spectrum) can be calculated by

\begin{equation}
\label{eq.6} P_{\rm{UCN}}(T_{\rm{n}})=N_{\rm{O_2}} \cdot
\int_{0}^{E ^{\rm{max}}} \frac {d\Phi (T_n)}{dE_0} dE_0 \cdot
\sigma _{UCN}(E_0).
\end{equation}

\begin{figure}[htp]
\includegraphics[width=0.5\textwidth]{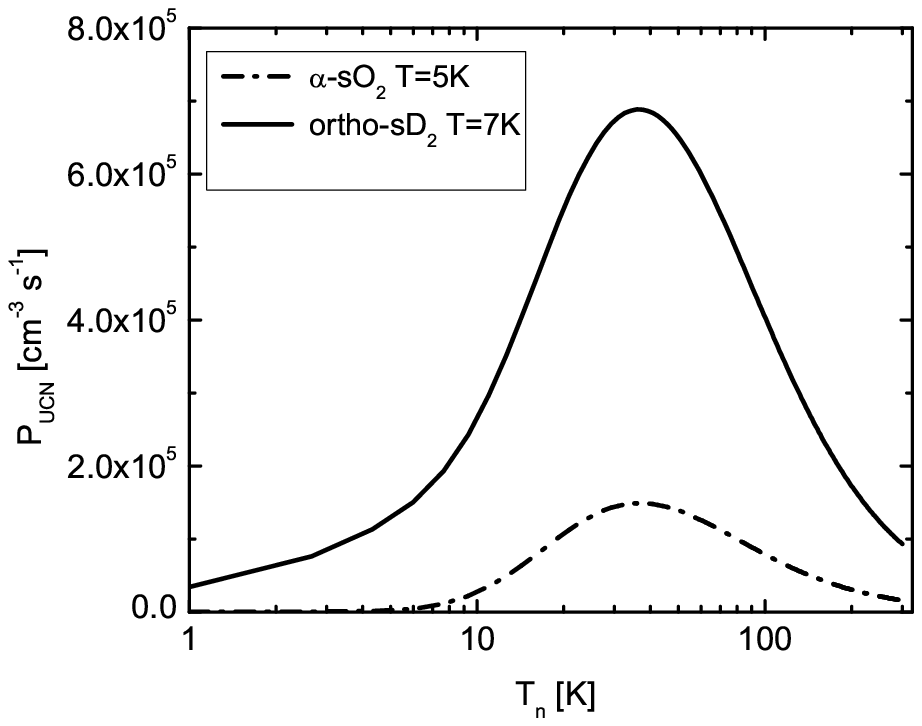}

\caption{~}{Calculated UCN production rate of ortho-sD$_2$ at
$T=$7~K (line) and $\alpha$-sO$_2$ at $T=$5~K (dash-dot line).
Both converters are exposed to a neutron capture flux of $\Phi
_C$=1$\cdot$10$^{14}$~cm$^{-2}$~s$^{-1}$} (Maxwellian shape with
effective temperature $T_n$). \label{fig.5}
\end{figure}

$N_{\rm{O_2}}=$~2.9$\cdot 10^{22}$~$cm^{-3}$ is the particle
density of the O$_2$ molecules. Fig.\ \ref{fig.5} shows the result
for $P_{UCN}(T_n)$ for sD$_2$ and sO$_2$. Contrary to the results
published in \cite{INDI} (optimal $T_{\rm{n}}\approx 10-15$~K) the
UCN production rate has a maximum at $T_{\rm{n}}\approx 40$~K.

In summary, new neutron scattering data of solid $\alpha$-oxygen
leads to a better understanding of UCN production in this
converter material. The new results  for the UCN production cross
section, resolved directly from the dynamical scattering function
$S(Q,E)$, show a significant UCN production cross section for
neutrons with energies at  $E_0\sim ~6$~meV. The leading
excitations are phonons and not magnons. This observation is
different to predictions of theoretical calculations \cite{INDI}
where the authors predict the contributions of magnons to the UCN
production, and identified these excitations as the leading
process of UCN production in solid $\alpha$-oxygen. An optimized
$\alpha$-sO$_2$ UCN source should be exposed to a cold neutron
flux with an effective neutron temperature of $T_n\simeq 40$~K,
where the production rate has a maximum. At this temperature the
UCN production rate of $\alpha$-sO$_2$ is only 22~\% of the
production rate of sD$_2$.

This work was supported by the cluster of excellence "Origin and
Structure of the Universe" (Exc 153) and by the
Maier-Leibnitz-Laboratorium (MLL) of the
Ludwig-Maximilians-Universit\"at (LMU) and the Technische
Universit\"at M\"unchen (TUM). We thank T. Deuschle, S. Materne ,
C. Morkel, and H. Ruhland for their help during the experiments.

\bibliography{sO2-UCNP-1-1}

\end{document}